\documentclass[reprint,
 amsmath,amssymb,
 aps,
 pra,
]{revtex4-2}
\usepackage[table,xcdraw]{xcolor}
\usepackage{graphicx}
\usepackage{dcolumn}
\usepackage{bm}
\usepackage{braket}
\usepackage{upgreek}
\usepackage[hidelinks,linkcolor=blue,colorlinks=true,urlcolor=blue,citecolor=blue]{hyperref}

\newcommand*\diff{\mathop{}\!\mathrm{d}}

\newcommand{\ee}{\mathrm{e}}
\newcommand{\ii}{\mathrm{i}}

\newcommand{\Gammat}{\tilde{\Gamma}}

\newcommand{\omegan}{\tilde{\omega}}

\newcommand{\acs}{{\underline{\mathrm{a}}}}
\newcommand{\GG}{\underline{\mathrm{G}}}
\newcommand{\dd}{\underline{\mathrm{d}}}
\newcommand{\Esp}{\underline{F}}

\newcommand{\Omegac}{\underline{\underline{\Omega}}}
\newcommand{\xx}{{\underline{\Psi}}}
\newcommand{\yy}{{\underline{\Phi}}}
\newcommand{\uu}{\underline{\psi}}
\newcommand{\vv}{\underline{\varphi}}
\newcommand{\uui}{\psi_i}
\newcommand{\vvi}{\varphi_i}
\newcommand{\uh}{\hat{\uu}}
\newcommand{\vh}{\hat{\vv}}
\newcommand{\Hmat}{\underline{\underline{H}}}
\newcommand{\II}{\underline{\underline{I}}}
\newcommand{\omegas}{\omega_s}

\newcommand{\inner}[2]{{\langle #1,#2\rangle}}

\renewcommand{\selectlanguage}[1]{}
\renewcommand{\Re}{\mathrm{Re}}
\renewcommand{\Im}{\mathrm{Im}}

\newcommand{\ddt}[1]{\frac{\diff {#1}}{\diff t}}
\begin{document}

\title{Instantaneous modes in dispersive laser cavities}

\author{Kristian Seegert}
\email{krsee@dtu.dk}
\author{Yi Yu}
\author{Mikkel Heuck}
\author{Jesper Mørk}
\affiliation{ Department of Electrical and Photonics Engineering, Technical University of Denmark, Ørsteds Plads 345A, 2800 Kgs. Lyngby, Denmark}
\affiliation{NanoPhoton - Center for Nanophotonics, Ørsteds Plads 345A, 2800 Kgs. Lyngby, Denmark}

\date{\today}

\begin{abstract}
We develop a unified instantaneous-mode description for lasers with dispersive cavities, exploiting the separation of timescales between fast cavity fields and slow carrier dynamics. The resulting reduced rate equations retain the essential effects of frequency-dependent mirrors through a dynamic modal gain and an effective confinement factor determined directly by the mirror reflectivity. Applied to a Fano laser, the reduced description captures the essential dynamics, closely reproduces the full-model behavior and clarifies the physical origin of dispersive instabilities. More generally, the approach provides a transparent framework for deriving reduced models and simplifying the stability analysis of dispersive laser cavities.
\end{abstract}

\maketitle

\section{\label{sec:level1}Introduction}

\begin{figure}
    \centering
    \includegraphics[width=1\linewidth]{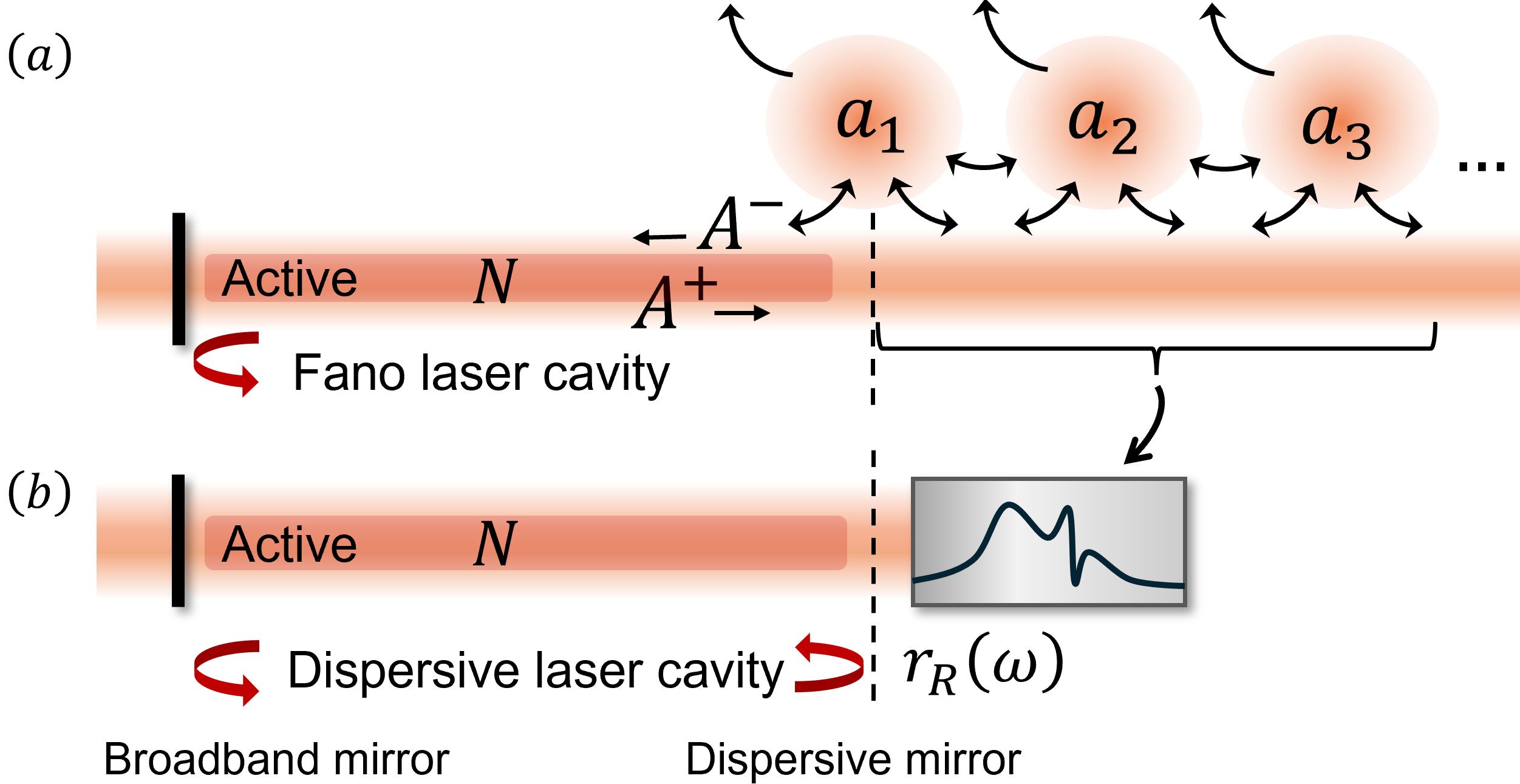}
    \caption{(a) Sketch of a Fano laser with $M$ side-coupled cavities. (b) Sketch of a dispersive cavity with one broadband (frequency-independent) mirror and one dispersive mirror. }
    \label{fig:laser_dispersive_mirror}
\end{figure}

Integrating a dispersive mirror in a semiconductor laser can lead to a wide range of complex dynamics. Dispersive laser cavities already play an important role as narrow-linewidth, tunable light-sources~\cite{komljenovic_widely_2015,corato-zanarella_widely_2023}, and also show rich dynamics, enabling engineering of various dynamical states, such as self-pulsing~\cite{feiste_18_1994,wenzel_mechanisms_1996,bandelow_correct_1996,renaudier_phase_2007,rimoldi_cw_2022,rimoldi_damping_2022,mak_linewidth_2019,seegert_self-pulsing_2024}, dual-mode lasing~\cite{mak_high_2021,seegert_self-pulsing_2024}, and chaos~\cite{mork_chaos_1992,radziunas_traveling_2024,rasmussen_suppression_2019}.

Since similar dynamical regimes occur across many devices, general analyses are valuable because they can distinguish universal effects from implementation-specific features when analyzing a particular dispersive laser. 

In this article, we present a general modal analysis of dispersive laser cavities based on an expansion in \textit{instantaneous modes}~\cite{piprek_multisection_2005, radziunas_traveling_2024,bandelow_correct_1996,seegert_self-pulsing_2024}. The instantaneous modes are the resonant modes (or the quasinormal modes~\cite{kristensen_modeling_2020}) of the dispersive laser cavity when the carrier density of the active medium is considered frozen at its present value; that is, the instantaneous modes solve the instantaneous eigenvalue problem. 

In the expansion, the mode amplitudes evolve dynamically via ordinary differential equations (ODEs), whereas the time dependence of the field distribution enters only parametrically through the instantaneous modes. This is particularly useful when only a small number of instantaneous modes are required to capture the relevant dynamics, thereby allowing for a straightforward dimensional reduction~\cite{bandelow_correct_1996,radziunas_traveling_2024}. The resulting models have reduced computational complexity, admit simpler analytical treatment, and provide greater transparency of the underlying physical mechanisms. 

This idea of expanding in instantaneous eigenstates is of course well known from the adiabatic theorem in quantum mechanics and condensed matter physics~\cite{born_beweis_1928,berry_quantal_1984}, and the Born-Oppenheimer approximation in quantum chemistry~\cite{born_zur_1927}. In lasers, instantaneous modes have previously been applied to chaotic external cavity lasers~\cite{radziunas_traveling_2024}, and to multi-section distributed feedback (DFB) lasers~\cite{wenzel_mechanisms_1996, bandelow_correct_1996}. 
In this article, we generalize to the broader class of lasers with dispersive mirrors [Fig.~\ref{fig:laser_dispersive_mirror}(b)], which includes the recent example of Fano lasers~\cite{mork_photonic_2014}  [Fig.~\ref{fig:laser_dispersive_mirror}(a)]. 

As a primary example, we consider a Fano laser with $M$ side-coupled cavities [Fig.~\ref{fig:laser_dispersive_mirror}(a)], which generalizes the conventional single-cavity Fano laser~\cite{mork_photonic_2014,mork_nanostructured_2025}. The conventional Fano laser leverages Fano interference between a continuum of waveguide modes and a discrete resonance from a side-coupled nanocavity~\cite{fano_effects_1961,fan_temporal_2003} to form a bound state in the continuum~\cite{yu_ultra-coherent_2021,mork_nanostructured_2025}. Many interesting features have already been demonstrated, including the possibility of terahertz range frequency modulation~\cite{mork_photonic_2014}, self-pulsing based on a saturable absorber~\cite{yu_demonstration_2017,rasmussen_theory_2017}, tolerance towards external feedback~\cite{rasmussen_suppression_2019}, ultra-narrow linewidth~\cite{yu_ultra-coherent_2021,yu_theory_2022}, optical bistability~\cite{liang_optical_2024}, and cavity-dumping by modulating the side-coupled cavity~\cite{dong_cavity_2023}. The extension to multiple cavities gives further possibilities to engineer the mirror dispersion towards self-pulsing and multi-mode lasing~\cite{seegert_self-pulsing_2024}.

As an important case, we show that a single-cavity Fano laser can, under appropriate conditions, develop undamped relaxation oscillations, and that this behavior is well captured by a single instantaneous mode. The instability mechanism can be identified as dispersive self-Q-switching, a form of self-pulsing also observed in DFB lasers~\cite{wenzel_mechanisms_1996,bandelow_dispersive_1996}, lasers with distributed Bragg-grating (DBR) mirrors~\cite{syvridis_large_1994,renaudier_phase_2007,tronciu_feedback_2021}, and coupled-cavity Fano lasers~\cite{seegert_self-pulsing_2024}. 

In the $M$-cavity Fano laser, the instability originates from a carrier-density-dependent redistribution of the intracavity field between the active region and the passive side-coupled cavities. Because the oscillation frequency depends on the carrier density through the linewidth-enhancement factor $\alpha$, variations in carrier density modify the field distribution and give rise to a dynamically varying longitudinal confinement factor. This dynamic confinement factor affects both the noise properties and the stability, and under certain conditions leads to undamped relaxation oscillations sustained by carrier-density-induced modulation of the confinement factor.

In the reduced single-mode equations, the dynamic confinement factor appears as a weighting of the stimulated-emission term and can be expressed entirely in terms of the effective reflectivity $r_R(\omega)$ and its derivative $\partial_\omega r_R(\omega)$. Crucially, this result is general: for dispersive laser cavities of the form sketched in Fig.~\ref{fig:laser_dispersive_mirror}(b), the ODEs governing the mode amplitudes depend only on $r_R(\omega)$, independent of the physical structure realizing the effective mirror. 

We first formulate the instantaneous-mode description for an $M$-cavity Fano laser within a transmission-line (ODE) model and derive reduced single-mode rate equations expressed directly in terms of the effective mirror reflectivity. We then validate the reduced model against full numerical simulations and use the reduced equations to analyze the onset of dispersive instabilities. Finally, we extend the approach to the traveling-wave model and demonstrate that the same reduced dynamics emerge in a fully distributed formulation.

 \section{Instantaneous modes in the ODE model}
We consider a transmission-line model~\cite{tromborg_transmission_1987} and specialize it to an $M$-cavity Fano laser, modeled using temporal coupled-mode theory~\cite{kristensen_theory_2017,wonjoo_suh_temporal_2004}. The ODE model is valid in a frequency range $|\omega-\omega_s|\tau_L\ll1$, where $\omega_s$ is the steady-state oscillation frequency and $\tau_L=2L/v_g$ is the round-trip time in the active section of length $L$ and group velocity $v_g$. This is generally satisfied for lasers with short active sections, but for lasers with longer active sections, the traveling-wave equation (TWE) model introduced later should be used. 
The dynamical variables are the slowly varying amplitude $A^+(t)$ of the incident field at the dispersive mirror, the nanocavity amplitudes $\acs(t) = (a_1,a_2,...,a_M)^T$, and the carrier density $N(t)$, as sketched in Fig.~\ref{fig:laser_dispersive_mirror}(a). The amplitudes are normalized such that $|A^+(t)|^2$ is the power at a reference plane just left of the dispersive mirror, and $|a_i(t)|^2$ is the energy in the $i$-th cavity.
The reflected field from the nanocavities is given by $A^-(t)= r_B A^+(t) +\dd^T\acs(t)$, where $\dd$ is a vector representing the coupling between waveguide and side-coupled cavities, in units of s$^{-\frac{1}{2}}$. Their Fourier transforms are related by $A^-(\omega)=r_R(\omega)A^+(\omega)$, where \begin{equation}
    r_R(\omega) = r_B +r(\omega),
\end{equation}
is the reflectivity of the dispersive mirror, with $r_B$ representing a broadband non-dispersive contribution and $r(\omega)$ being the dispersive part. 

The dynamical model takes the following form,

\begin{align}
    \ddt{A^+}&=\frac{1}{2}(1-\ii\alpha)\left(\Gamma v_gg(N)-\frac{1}{\tau_p}\right)A^+\nonumber\\
    &+\gamma_L\left(\frac{r_B A^+ +\dd^T\acs}{r_R(\omega_s)}-A^+\right)+F_A(t), \label{eq:ode1}\\
    \ddt{\acs}&=-\ii \Omegac\,\acs+\dd A^+,\\
    \ddt{N} &= R_p-\frac{N}{\tau_s}-v_gg(N)N_p,\label{eq:ode3}
\end{align}
where $\alpha$ is the linewidth enhancement factor, $\Gamma$ is the confinement factor, $v_g$ is the group index, $g(N)$ is the gain, $\tau_p$ is the photon lifetime, $\gamma_L=\frac{1}{\tau_L}$ is the inverse round-trip time in the active section, and $F_A(t)$ is a zero-mean Langevin force corresponding to spontaneous emission noise. 
The matrix $\Omegac$ is a non-Hermitian, but symmetric matrix representing the modes of the individual nanocavities and the coupling between them. Finally, $R_p$ is the pump rate, $\tau_s$ is the carrier lifetime, and $N_p$ is the photon number density in the active section, which in steady-state is proportional to $N_p=C(\omega_s,N_s)|A^+|^2$, and in the ODE model, this proportionality is assumed to hold out of equilibrium also. 

The reflected field $A^-(t)$ can additionally be related to the incident field $A^+(t)$ through the impulse response of the dispersive mirror, \begin{equation}
    \label{eq:impulse}
    A^-(t)=r_BA^+(t)+\int_{0}^\infty r(\Delta t)A^+(t-\Delta t)\diff \Delta t,
\end{equation} where $r(\Delta t)$ is the Fourier transform of $r(\omega)$.

Note that the dimensionality of the full ODE system is $D=1+2(M+1)$, corresponding to a (real) carrier density, and $M+1$ complex envelopes. 

Now, the key to the modal approach is to realize that the equations governing the dynamics of $A^+$ and $\acs$ are linear when $N$ is interpreted as a time-dependent parameter. The full system describing the cavity fields can then be written as,
\begin{equation}
\label{eq:HamiltonianODE}
\ddt{\xx}=-\ii \Hmat(N)\xx+\underline{F}_A, 
\end{equation}
where $\xx=(A^+,\acs^T)^T$, $\underline{F}_A(t)=(F_A(t),\underline{0}^T)^T$, and $\Hmat$ is given in the Appendix. The idea is to expand $\xx$ in its instantaneous modes. We define the instantaneous modes of the ODE model as solutions to the instantaneous time-independent eigenvalue problem for a frozen $N$,
\begin{equation}
    \Hmat(N)\uu_n(N) =\omegan_n(N) \uu_n(N),
\end{equation}
where $\uu_n(N)=(\psi_n^{(A)},\psi_n^{(a_1)},...,\psi_n^{(a_M)})^T$ is a right-eigenvector and $\omegan_n=\omega_n-\ii\gamma_n$ is the instantaneous mode frequency, with $\omega_n$ representing a resonance frequency and $\gamma_n$ the corresponding net decay rate. Equivalently, the imaginary part may be expressed as the effective net modal gain per unit time, $G_n=2\Im(\omegan_n)$.

Since $\Hmat(N)$ is non-Hermitian, the right-eigenvectors are not generally orthogonal $\inner{\uu_m}{\uu_n}\neq0$ under the usual inner-product $\inner{\vv}{\uu}=\sum \vvi^*\uui$. Instead, the right-eigenvectors are biorthogonal to a set of adjoint modes that solve $\Hmat(N)^\dagger \vv_n=\omegan_n^*\vv_n$, where $\vv_n=(\varphi_n^{(A)},\varphi_n^{(a_1)},...,\varphi_n^{(a_M)})^T$ are the left-eigenvectors~\cite{brody_biorthogonal_2014}. 
The eigenvalues of the adjoint problem are the complex conjugates of the original problem. The right- and left-eigenvectors can then be normalized in the bi-orthogonal sense
\begin{equation}
    \inner{\vh_m}{\uh_n}=\delta_{mn},
\end{equation}
which still leaves one degree of freedom for defining the normalized eigenvectors. We now expand \begin{equation}
    \xx(t) =\sum_mf_m(t)\uh_m[N(t)].
\end{equation}
Inserting in (\ref{eq:HamiltonianODE}) results in, 

\begin{equation}
\label{eq:fn}
    \ddt{f_n}=-\ii\omegan_nf_n-\sum_m\inner{\vh_n}{\partial_t{\uh_m}}f_m+F_n,
\end{equation}
where $F_n(t)=\inner{\vh_n}{\underline{F}_A}$. Evidently, the different modes are coupled through the time derivative of the eigenvectors $\inner{\vh_n}{\partial_t{\uh_m}}=\inner{\vh_n}{\partial_N{\uh_m}}\ddt{N}$. In general, the term $\inner{\vh_n}{\partial_t\uh_n}$ also appears in (\ref{eq:fn}), which gives rise to a geometric phase~\cite{berry_quantal_1984}; however, we choose a gauge where this term vanishes, which are the so-called parallel transported eigenvectors~\cite{ibanez_adiabaticity_2014}. With this normalization, we have \begin{equation}
    A^+(t)=\sum_n\sqrt{\Gammat_n}f_n(t),
\end{equation}
where $\Gammat_n$ is a complex-valued dynamically varying longitudinal confinement factor defined by \begin{equation}
    \Gammat_n = \frac{\inner{\vv_n}{\uu_n}_A}{\inner{\vv_n}{\uu_n}},\qquad \inner{\vv_n}{\uu_n}_A=\varphi_n^{(A)}\psi_n^{(A)}.
\end{equation}

It can be written solely in terms of the dispersive mirror reflectivity and its derivative
\begin{equation}
\label{eq:Gamma-ode}
    \Gammat_n=\left(1-\ii\gamma_L\frac{\partial_\omega r_R(\omegan_n)}{r_R(\omega_s)}\right)^{-1}.
\end{equation}
This result holds in general for an arbitrary reflectivity. The complex confinement factor can be related to the derivative of the instantaneous mode frequency with respect to the carrier density \begin{equation}
    \partial_N\omegan_n=\frac{\inner{\vv_n}{(\partial_N\Hmat)\uu_n}}{\inner{\vv_n}{\uu_n}}=\Gammat_n\beta_N,
\end{equation} where $\beta_N=\frac{1}{2}(\ii+\alpha)\Gamma v_g g_N$. 
The coupling terms for $m\neq n$ can be evaluated as 
\begin{equation}
\label{eq:coupling}
    \inner{\vh_n}{\partial_t{\uh_m}}=\frac{\inner{\vh_n}{\partial_t\Hmat\,{\uh_m}}}{\omegan_m-\omegan_n}=\frac{\sqrt{\Gammat_m\Gammat_n}\beta_N}{\omegan_m-\omegan_n}\ddt{N}.
\end{equation}

From (\ref{eq:coupling}), it is clear that this transformation breaks down at exceptional points, where two or more eigenvalues and their eigenvectors coalesce~\cite{ozdemir_paritytime_2019}. While the modal approach could possibly be modified to incorporate this using Jordan-decomposition, it is unclear whether there is a significant advantage to be gained here compared to maintaining a fixed basis. As exceptional points are not the focus of this paper, we leave aside this special case.

So far, the approach is exact. However, it is now straightforward to consider the case where a single mode $\omegan_n$ is dominant, i.e., all other modes have much larger effective decay rates $\gamma_m\gg \gamma_n$ for $n\neq m$. Defining $f_n=\sqrt{S_n/C(N_s,\omega_s)}\exp(\ii\phi_n)$, we finally obtain reduced single-mode equations for the ODE model
 \begin{align}
    \label{eq:ode-reduced-S}
    \ddt{S_n}&=G_n(N)S_n+R_n(N)+F_{Sn},\\
    \label{eq:ode-reduced-phi}
    \ddt{\phi_n}&=-\omega_n(N)+F_{\phi n},\\
     \label{eq:ode-reduced-N}
    \ddt{N}&=R_p-\frac{N}{\tau_s}-v_gg(N)|\Gammat_n(N)|S_n,
\end{align}
where $R_n(N)$ is the mean spontaneous emission rate into the $n$'th mode, and $F_{Sn}$ and $F_{\phi n}$ are zero-mean Langevin noise terms. The correlations for these are given in the Appendix for the more general TWE model considered later. Note that we have omitted the carrier noise term, which is not important for the intrinsic linewidth~\cite{henry_theory_1986}. 

In the end, the dynamical system has been reduced to a two-dimensional system (the phase equation is decoupled) from the original $1+2(M+1)$ dimensions, leading to drastically simpler analysis. Compared to the conventional rate equations for non-dispersive cavities, the gain term in the photon equation has been replaced by the effective modal gain of the instantaneous mode, which resides partly outside the active section. In addition, the stimulated emission term in the carrier density equation is weighted by the absolute value of the complex confinement factor $|\Gammat_n(N)|$.

\begin{figure}
    \centering
    \includegraphics[width=\linewidth]{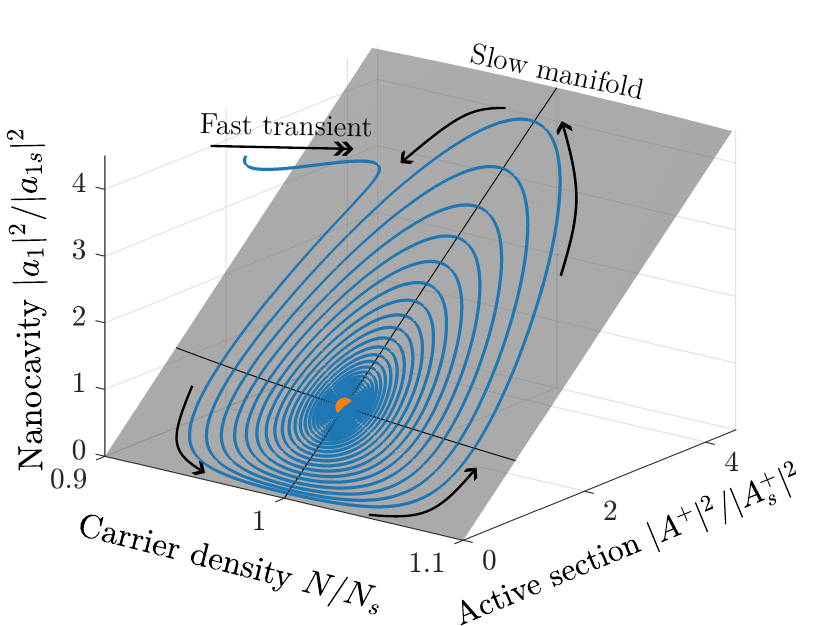}
    \caption{A trajectory in phase-space $(N,|A^+|^2,|a_1|^2)$ for the Fano laser. The slow manifold is marked in grey, the orange dot shows the steady-state point, the arrows indicate the flow on the slow manifold, and the black lines indicate the nullclines. }
    \label{fig:center_manifold_stable}
\end{figure}
\begin{figure}
    \centering
    \includegraphics[width=\linewidth]{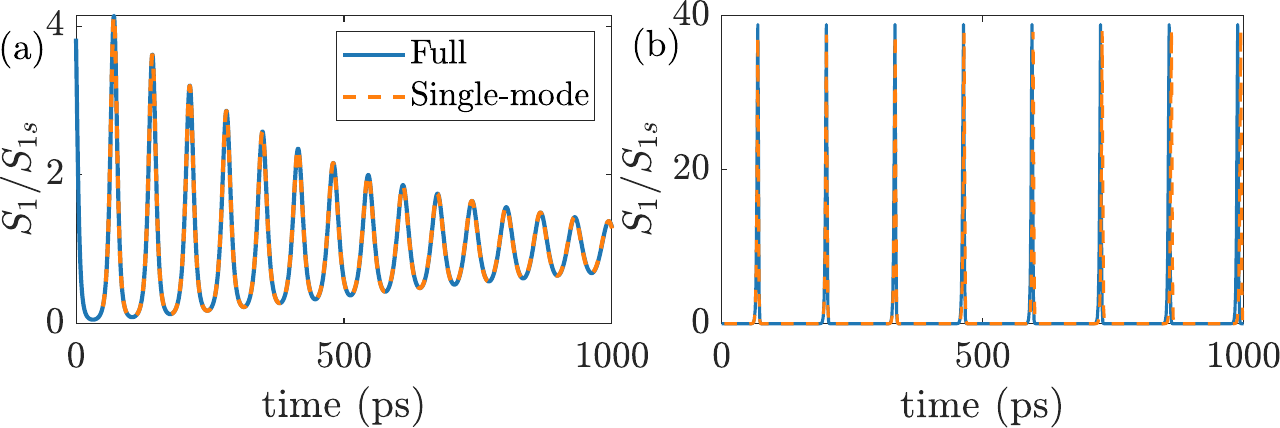}
    \caption{Timetraces of $S_1$ comparing the full ODE model (continuous blue line) with the single-mode approximation (dashed orange line). (a) Parameters leading to damped relaxation oscillations. (b) Parameters leading to a self-pulsing regime.}
    \label{fig:single-mode}
\end{figure}
\begin{figure}
    \centering
    \includegraphics[width=\linewidth]{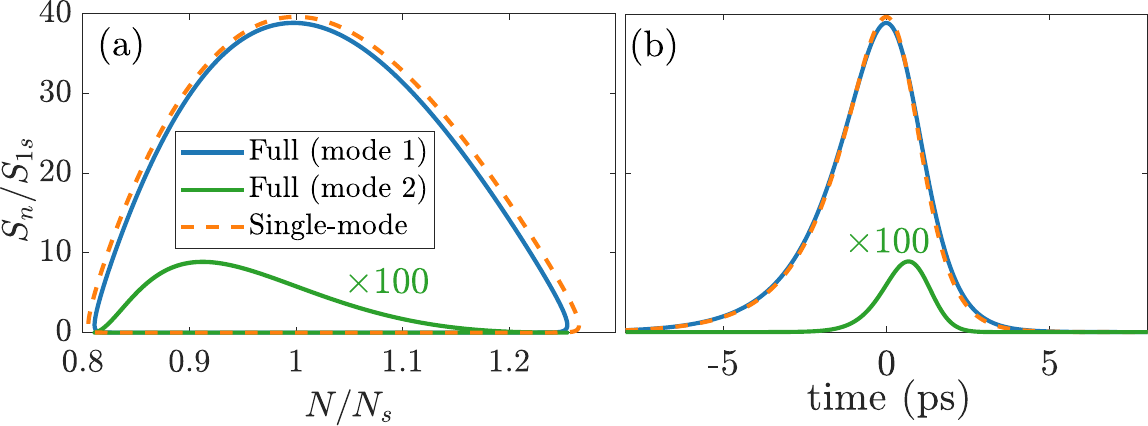}
    \caption{(a) Limit-cycle projected onto $(N,S_n)$-plane for mode 1 (blue) and 2 (green), together with the single-mode approximation of mode 1. (b) Corresponding timetrace of a single pulse.}
    \label{fig:self-pulsing}
\end{figure}

As an example, we consider the conventional Fano laser with a single side-coupled cavity. For these simulations, we ignore the noise terms and spontaneous emission into the lasing mode. We use the parameters from Ref.~\cite{yu_ultra-coherent_2021}, but set $\alpha=5$, $\tau_s =0.8$ ns and $L=15.3\upmu$m to illustrate richer dynamics. With these parameters, we have $\tau_L=0.32$ ps, and the range of validity of the ODE model is $|\omega-\omega_s|\ll2\pi\times490$ GHz, which should be satisfied.

In Fig.~\ref{fig:center_manifold_stable}, we show the results of a typical simulation for a pump rate $R_p=4N_s/\tau_s$, and with $\omega_s-\omega_c=-\gamma_t$, where $\omega_c$ is the nanocavity resonance frequency and $\gamma_t$ is the total decay rate of the cavity. The trajectory is projected onto the three-dimensional phase space $(N,|A^+|^2,|a_1|^2)$, with axes normalized to their values at steady state. 
We observe that the trajectory is quickly attracted to a slow manifold shown in grey, and subsequently undergoes relaxation oscillations toward a steady state marked by the orange dot. The slow manifold is the surface spanned by $(N,S_1)\rightarrow (N,S_1|\psi^{(A)}_{1}(N)|^2,S_1|\psi^{(a_1)}_{1}(N)|^2)$, where the subscript $n=1$ refers to the lowest threshold mode. The black lines indicate the nullclines on the slow manifold $\ddt{N}=0$ and $\ddt{S}=0$. 
The plot illustrates the slow-fast nature of the system, where non-dominant modes quickly die out.

Figure~\ref{fig:single-mode} shows the timetraces of a typical simulation for two different sets of parameters ($\omega_s-\omega_c=-\gamma_t$ and $\omega_s-\omega_c=-1.51\gamma_t$), leading to respectively damped (a) and undamped (b) relaxation oscillations. The undamped relaxation oscillations develop into periodic self-pulsations, shown in the figure. The y-axis is the projection onto the $n=1$ mode ($S_1=|\inner{\vh_1(t)}{\xx(t)}|^2$), normalized to its value at steady state. For comparison, the reduced single-mode equations were simulated and plotted on top (orange dashed line) with initial conditions corresponding to the full calculation after a small delay. We observe excellent agreement, and, importantly, the reduced equations also capture the dynamics in the self-pulsing regime. 

Figure~\ref{fig:self-pulsing}(a) shows the limit-cycle projected onto the $(N,S_n)$-plane for mode 1 (blue) and mode 2 (green) together with the single-mode approximation (orange dashed). A zoom-in on the timetrace of a single pulse is shown in Fig.~\ref{fig:self-pulsing}(b). The amplitude of mode 2 remains very small and has been scaled by a factor of 100 for visibility in both plots. We observe that the limit-cycle in the single-mode approximation is slightly larger, but displays the same dynamics as the full calculation, thus demonstrating the single-mode nature of the pulsing. The small deviation is attributed to the fast variation of the carrier density during the pulse, which introduces a small non-adiabatic coupling to mode 2. 

We emphasize that the ability to capture the dynamics far from steady state distinguishes the approach from the steady-state ab initio laser theory (SALT)~\cite{tureci_self-consistent_2006}.

\section{Stability analysis}
The instability leading to the self-pulsing regime in Fig.~\ref{fig:single-mode}(b) can be studied using the reduced equations, drastically simplifying the stability analysis.  

The eigenvalues of the linearized system are given by\begin{equation}
    \lambda=-\gamma_R\pm \ii \sqrt{\omega_R^2 -\gamma_R^2},
\end{equation}
where the damping rate is 
\begin{equation}
\label{eq:gammaNN}
    \gamma_{R} = \frac{1}{2\tau_s}+\frac{1}{2}v_g \left(|\Gammat_{n}|_N g_{th}+|\Gammat_n|g_N\right)S_s,
\end{equation}
and the relaxation resonance frequency given by \begin{equation}
    \omega_R^2 = v_g g_{th}G_{nN}|\Gammat_n|S_s.
\end{equation}
The subscript $N$ denotes derivative with respect to the carrier density, and all the parameters $g_N$, $G_{nN}$, $|\Gammat_n|$, and $|\Gammat_n|_N$ are evaluated at the steady-state carrier density $N_s$. 
Note that if we replace $|\Gammat_n|G_{nN}$ with $\Gamma v_g g_N$ we get the expression for a Fabry-Pérot cavity. Second, note that $\gamma_{R}$ can become negative if and only if the parenthesis in (\ref{eq:gammaNN}) is negative, which further requires $|\Gammat_n|_N<0$. That is, the system can become unstable if increasing the carrier density pushes the field distribution out of the active section, such that an increase in carrier density \textit{reduces} the stimulated recombination. This is essentially the same conclusion drawn in Ref.~\cite{chun-lin_single_2003} when analyzing an asymmetrically pumped multi-section DFB laser; however, here it is generalized to an arbitrary reflectivity. 

Finally, we note that the noise spectrum is easily obtained in the single-mode approximation using conventional linewidth theory~\cite{henry_theory_1986,coldren_diode_2012,tromborg_theory_1991}. The end result is \begin{equation}
    \Delta\nu =\frac{R_n}{4\pi S_n}(1+\overline{\alpha}^2)=\Delta\nu_{0}|\Gammat_n|^2(1+\overline{\alpha}^2),
\end{equation}
where $\Delta\nu_{0}$ is the bare Schawlow-Townes linewidth of an equivalent laser with a frequency-independent right mirror $r_2=r_R(\omega_s)$, and $\overline{\alpha}=\partial_N\Re(\omegan_n)/\partial_N\Im(\omegan_n)$ is the effective linewidth enhancement factor. Evidently, the linewidth is increased due to the reduced confinement in the active section, and the broadening due to phase-amplitude coupling is treated with the effective linewidth enhancement factor. Further, letting $r_R(\omega)=\exp[\rho_R(\omega)+\ii\phi_R(\omega)]$, one can rewrite $|\Gammat_n|^2(1+\overline{\alpha}^2)=(1+\alpha^2)/F^2$, where $F=1+\partial_\omega \phi_R/\tau_L+\alpha\partial_\omega \rho_R/\tau_L$ is the so-called chirp-reduction factor~\cite{tromborg_transmission_1987,yariv_self-quenching_1990}. Thus, the single-mode approximation also correctly captures well-known results regarding linewidth reduction. 
\section{Instantaneous modes in the TWE model}
While the ODE model is itself already an approximate model, the ideas and conclusions are straightforward to generalize. To illustrate this, we consider the Traveling Wave Equations (TWE)~\cite{tromborg_traveling_1994,bandelow_correct_1996,wenzel_semiconductor_2021,radziunas_traveling_2024}. 
The TWE model is a PDE, 

\begin{align}
\ii \partial_t \xx &= \Hmat(z,t) \xx(z,t)+\underline{F}(z,t),\quad \xx(z,t)=\begin{pmatrix}
A^+(z,t) \\
A^-(z,t)
\end{pmatrix},
\end{align}
where $\underline{F}(z,t)$ is a distributed Langevin force term and
\begin{equation}
    \Hmat(z,t) = v_g
\begin{pmatrix}
- \ii\partial_z - \beta(t) & 0 \\
0 & \ii\partial_z - \beta(t)
\end{pmatrix},
\end{equation}
with $\beta(t) = -\frac{1}{2}(\ii+\alpha)\Gamma g[N(t)]+\frac{1}{2}\ii\alpha_i$. $\xx(z,t)$ satisfies the boundary conditions $A^+(-L,t)=r_1A^-(-L,t)$ and $A^-(0,\omega)=r_R(\omega)A^+(0,\omega)$, such that the relation between $A^-(0,t)$ and $A^+(0,t)$ is given by the impulse response in (\ref{eq:impulse}). We assume $\beta(t)$ to be constant within the active region, thereby neglecting spatial hole burning. The instantaneous modes of the TWE model are the solutions $\Hmat\,\uh=\omegan\uh$ satisfying the boundary conditions. 

In the Appendix, we show how to apply the modal approach using a trick to turn the non-local boundary conditions of the dispersive mirror into a virtual extended cavity with local boundary conditions that are straightforward to handle. In the end, if the $n$'th mode is dominant, then the dynamics can again be reduced to a two-dimensional system governed by (\ref{eq:ode-reduced-S})--(\ref{eq:ode-reduced-N}), provided that we substitute $|\Gammat^{(ode)}_n|\rightarrow \sqrt{K_{n}}|\Gammat^{(twe)}_n|$. Here, the complex confinement factor is \begin{equation}
\label{eq:Gamma_mm}
    \Gammat_n=\frac{\tau_L}{\tau_L+\tau_R(\omegan_n)}\qquad \textrm{(TWE model)},
\end{equation}
and $ \tau_R(\omega)=-\ii\partial_\omega \ln r_R(\omega)$ is a complex round-trip time in the dispersive mirror, and \begin{equation}
\label{eq:sigma}
    K_n=\left|\frac{\left(|r_1|+|r_R(\omegan_n)|\right)(1-|r_1r_R(\omegan_n)|)}{(-\ln|r_1r_R(\omegan_n)|^2)|r_1r_R(\omegan_n)|}\right|^2,
\end{equation}
is the longitudinal Petermann excess spontaneous emission factor~\cite{petermann_calculated_1979} defined over the active section only.
Equation (\ref{eq:Gamma_mm}) is similar to Eq.~(\ref{eq:Gamma-ode}), but with $r_R(\omega_s)$ exchanged for $r_R(\omegan)$, which reflects the fact that the TWE model is valid over a wider frequency range. Since the reduced equations of the TWE model are formally equivalent, the small-signal analysis carries over. 

\section{Conclusion}

We have introduced an instantaneous-mode framework for dispersive laser cavities, enabling a physically transparent interpretation of their dynamics through the effective mirror reflectivity. Beyond enabling reduced models, the approach offers a diagnostic perspective on laser instabilities by revealing whether the essential dynamics are governed by a single mode or require multimode interactions. This distinction is particularly relevant for interpreting self-pulsing, multimode operation, and the onset of complex dynamics in dispersion-engineered cavities. By reducing the dynamics to low-dimensional ODEs in relevant regimes, the framework also makes it natural to apply bifurcation theory to relaxation oscillations, including the possible emergence of global bifurcations such as homoclinic orbits.
\begin{acknowledgments}
This work was supported by the Danish National Research Foundation (Grant No. DNRF147 - NanoPhoton). M.H. and Y.Y. acknowledge the support from Villum Foundation via the Young Investigator Programme (Grant No. 37417 - QNET-NODES, and Grant No. 42026 - EXTREME).
\end{acknowledgments}
\section*{Data availability statement}
The data that support the findings of this article are openly available~\cite{seegertDATASETInstantaneousModes2026}.
\appendix

\section{Instantaneous modes in the ODE model}
\label{app:instantaneousODE}
The $\Hmat$ and $\xx$ in (\ref{eq:HamiltonianODE}) are given by, 
\begin{equation}
\label{eq:xH}
    \xx=\begin{pmatrix}
        A^+\\
        \acs
    \end{pmatrix}, \qquad \Hmat(N) = \begin{pmatrix}
        \ii\gamma_1(N)&\ii\gamma_2 \dd^T\\
        \ii\dd &\Omegac
    \end{pmatrix},
\end{equation}
with \begin{align}
    \gamma_1(N)&=\Delta G(N)-\gamma_L\left(1-\frac{r_B}{r_R(\omegas)}\right),&
    \gamma_2&=\frac{\gamma_L}{r_R(\omegas)}.
\end{align}
Here, $\Delta G(N)=\frac{1}{2}(1-\ii\alpha)\left(\Gamma v_gg(N)-\frac{1}{\tau_p}\right)$. The instantaneous mode frequencies of the ODE model satisfy
\begin{equation}
    -\ii\omegan=\gamma_1(N)+\gamma_2r(\omegan).
\end{equation}
A corresponding right-eigenvector to the eigenfrequency $\omegan_n$ is given by 
\begin{equation}
    \uu_{n}=\begin{pmatrix}
        1\\
        \GG(\omegan_n)
    \end{pmatrix},
\end{equation}
where $\GG(\omega)$ relates the amplitudes $\acs(\omega)$ and $A^+(\omega)$, $\acs(\omega)=\GG(\omega)A^+(\omega)$. It is given by \begin{equation}
    \GG(\omega)=\frac{\dd}{\ii(\Omegac-\II\omega)}.
\end{equation}
For the left-eigenvectors, it is more convenient to work with their complex conjugates. They can be written as
\begin{equation}
    \vv_n^*=\begin{pmatrix}
        1\\
        \gamma_2\GG(\omegan_n)
    \end{pmatrix}.
\end{equation}
Requiring that $\inner{\vv_n}{\uu_n}=1$ gives
\begin{equation}
\label{eq:uhvh}
    \uh_n=\frac{k_n\uu_n}{\sqrt{1+\gamma_2\GG(\omegan_n)^2}}, \quad \vh_n^* =\frac{\vv_n^*}{k_n\sqrt{1+\gamma_2\GG(\omegan_n)^2}},
\end{equation}
where $k_n$ is an arbitrary function of $N$ and $\GG(\omega)^2\equiv\GG(\omega)\cdot \GG(\omega)$. Requiring further that $\inner{\vv_n}{\partial_t\uu_n}=0$ means that $k_n$ must be constant, and we can set it to 1. If $\xx(t)=f_n(t)\uh_n(t)$, the energy in the active section in a single mode is proportional to $|\Gammat_n||f_n|^2$, where
\begin{equation}
    \Gammat_n=\left(1+\frac{\gamma_L}{r_{R}(\omegas)}\GG(\omegan_n)^2\right)^{-1}.
\end{equation}
Finally, we can rewrite this in terms of the reflectivity $r_R(\omegan)$ as
 \begin{equation}
    \GG(\omegan_n)^2=\dd^T\left(\frac{1}{\ii(\Omegac-\II\omegan_n)}\right)^2\dd =-\ii\partial_\omega r_R(\omegan_n).
\end{equation}
We then get the expression of $\Gammat_n$ for the ODE model. 
\section{Instantaneous modes in the TWE model}
\label{app:instantaneousTWE}
In the TWE model the instantaneous modes are given in the active section by \begin{equation}
    \uu_n(z)=C_n\begin{pmatrix}
        r_1\ee^{\ii(\tilde{k}_n+\beta)(z+L)}\\
        \ee^{-\ii(\tilde{k}_n+\beta)(z+L)}
    \end{pmatrix},\qquad z\leq0,
\end{equation}
where $\tilde{k}_n=\omegan_n/v_g$ satisfies the oscillation condition\begin{equation}
    r_1r_R(\omegan_n)\ee^{2\ii (\tilde{k}_n+\beta)L}=1.
\end{equation}
The nonlocal boundary conditions given by (\ref{eq:impulse}) can be handled by treating the dispersive reflector as a passive section with a Hamiltonian 
\begin{equation}
    \Hmat(z>0,t) = v_g\begin{pmatrix}
        -\ii\partial_z&0\\
        \kappa(z)&\ii\partial_z
    \end{pmatrix},
\end{equation}
where 
\begin{equation}
    \kappa(z)=\frac{2\ii}{v_g}r\left(\frac{2z}{v_g}\right)
\end{equation}
is a one-way coupling determined by the impulse response. The left-propagating fields on each side of the reference plane at $z=0$ are then related by $A^-(0^-,t)=A^-(0^+,t)+r_BA^+(0,t)$. Further, we demand that $A^-(z,t)\rightarrow 0$ for $z\rightarrow \infty$. One can show that if this is satisfied, we have 
\begin{align}
     A^+(z>0,t)&=A^+(0,t-z/{v_g}),\\
    A^-(z>0,t)&=\int_{2z/v_g}^\infty r(\Delta t)A^+(0,t+z/v_g-\Delta t)\diff \Delta t. 
\end{align}
This gives the correct expression for $A^-(0^+,t)$. The instantaneous modes are then given in the virtual extended cavity by 
\begin{equation} 
\uu_n(z) = D_n\begin{pmatrix}
    \ee^{\ii \tilde{k}_nz}\\
    \ee^{-\ii\tilde{k}_nz}\int_{2z/v_g}^\infty r(\Delta t)\ee^{\ii\omegan_n\Delta t}\diff \Delta t
\end{pmatrix}, \quad z>0,
\end{equation}
where  $D_n=C_nr_1\ee^{\ii\tilde{k}_nL}$. 
Now, to define a projection, it turns out to be more convenient to consider the scalar product \begin{equation}
    (\yy,\xx)=\int_{-L}^{\infty}\yy\cdot \xx \diff z. 
\end{equation}
We consider the adjoint $\Hmat^\ddagger$, which satisfies $(\yy,\Hmat\,\xx)=(\Hmat^\ddagger\yy,\xx)+B$, where $B$ is a boundary term. The adjoint problem is defined as \begin{equation}
    \Hmat^\ddagger\vv=\omegan^\ddagger\vv_,
\end{equation}
where $\vv$ is chosen to satisfy boundary conditions such that the boundary term $B$ vanishes in $(\vv,\Hmat\,\uu)=(\Hmat^\ddagger \vv,\uu)$ if the eigenvalues are identical $\omegan^\ddagger=\omegan$. In the end, one finds that the adjoint problem is identical to the original problem, if we simply reverse the direction of propagation. The adjoint eigenvalues will be the same as the original eigenvalues $\omegan^\ddagger_n=\omegan_n$. Thus, the adjoint modes are given by 
\begin{equation}
    \vv_n(z)=E_n \begin{pmatrix}
        \psi_n^-(z)\\
        \psi_n^+(z)
    \end{pmatrix}
\end{equation}
It turns out that the boundary term vanishes for all $m$ and $n$, such that $(\vv_m,\Hmat\,\uu_n)=(\Hmat^\ddagger\vv_m,\uu_n)$. Therefore, $(\omegan_n-\omegan_m)(\vv_m,\uu_n)=0$. We normalize $(\vv_n,\uu_n)=1$. 
Now, we expand \begin{equation}
    \xx(z,t)=\sum_nf_n(t)\uu_n(z,\beta(t)).
\end{equation}
Taking the time-derivative and projecting with $(\vv_n,\cdot)$ gives
\begin{equation}
    \ddt{f_n}=-i\omegan_nf_n-\sum_m(\vv_n,\partial_t \uu_m)f_m.
\end{equation}
We note that $\partial_t\uu_n=\partial_\beta\uu_n \ddt{\beta}$. Further, by looking at the derivative $\partial_\beta (\Hmat\,\uu_n)=\partial_\beta(\omegan_n\uu_n)$, one can show the following two identities
\begin{align}
    \partial_\beta \tilde{\omega}_n &= (\vv_n, (\partial_\beta \Hmat)  \uu_n ),\\
    (\vv_n,\partial_\beta \uu_m)&=\frac{(\vv_n,(\partial_\beta \Hmat)\uu_m)}{\omegan_m-\omegan_n},\qquad m\neq n. 
\end{align}
Now, in the active section $\partial_\beta\Hmat=-v_g\II$, while in the passive section $\partial_\beta \Hmat=0$. At the same time, the derivative $\partial_\beta \omegan$ can be calculated from the oscillation condition to give
\begin{equation}
    \partial_\beta\omegan_n=-v_g\Gammat_n,\qquad \Gammat_n=\frac{\tau_L}{\tau_L+\tau_R(\omegan_n)},
\end{equation}
with $\tau_R(\omegan)=-\ii\partial_\omega \ln(r_R(\omegan))$. This means that 
\begin{equation}
    (\vv_n,\uu_n)_A=\Gammat_n,
\end{equation}
where $(\cdot,\cdot)_A$ means the integral is over the active section only. The integral over the active section can be straightforwardly calculated as $(\vv_n,\uu_n)_A=2r_1LC_n^2E_n$, which gives a relation between $C_n$ and $E_n$. If we further require that $(\vv_n,\partial_\beta \uu_n)=0$, we must have $E_n$ constant, and we set $E_n=1$. Then, \begin{equation}
    C_n=\sqrt{\frac{\Gammat_n}{2r_1L}}.
\end{equation} 
In the carrier density equation, the stimulated emission term is 
\begin{align}
    R_{st}&=\frac{\Gamma v_gg(N)}{\hbar \omega V_A}\frac{1}{v_g}(\xx^*,\xx)_A,\\
    &=\frac{\Gamma v_gg(N)}{\hbar \omega V_A}\frac{1}{v_g}\sum_{n, m}f_n^*f_m(\uu_n^*,\uu_m)_A,\\
    &\label{eq:single-mode-TWE}\simeq \frac{\Gamma v_gg(N)}{\hbar \omega V_A}\frac{1}{v_g}|f_n|^2(\uu_n^*,\uu_n)_A,
\end{align}where (\ref{eq:single-mode-TWE}) is for a single dominant mode. Defining \begin{equation}
    S_n=\frac{\Gamma}{\hbar\omega V_Av_g}|f_n|^2,\quad K_{n}=\frac{(\vv_n^*,\vv_n)_A(\uu_n^*,\uu_n)_A}{|(\vv_n,\uu_n)_A|^2},
\end{equation}
we can write $(\uu_n^*,\uu_n)_A=|\Gammat_n|\sqrt{K_n}$, and 
we finally obtain \begin{equation}
    R_{st}= v_gg(N)\sqrt{K_n}|\Gammat_n|S_n,
\end{equation}
with $\sqrt{K_n}$ given in (\ref{eq:sigma}).

\section{Noise}
We assume uncorrelated white Langevin noise with second-order correlation functions $    \langle \Esp(z,t)\Esp^T(z',t')\rangle=\langle \Esp(z,t)^*\Esp^\dagger(z',t')\rangle=\underline{\underline{0}}$, and
\begin{equation}
    \langle \Esp(z,t)\Esp^\dagger(z',t')\rangle=2D_{FF}(z,t)\II\delta(z-z')\delta(t-t'),
\end{equation}
where $D_{FF}$ is noise correlation strength given by~\cite{henry_theory_1986}
\begin{equation}
    D_{FF}(z,t)=Q g[N(z,t)]n_{sp}[N(z,t)],
\end{equation}
where $n_{sp}$ is the population inversion factor, and $Q$ is a constant. Since we assume constant carrier density in the active section, $D_{FF}(z,t)=D_{FF}(t)$. The spontaneous emission that gets picked up by the $n$'th mode is $F_n(t)=(\vv_n,\Esp)_A$. The correlation functions for $F_n$ are
\begin{equation}
    \langle F_n(t)F_m^*(t')\rangle=2D_{nm}(t)\delta(t-t'),
\end{equation}
where $D_{nm}(t)=D_{FF}(t)(\vv_n,\vv_m^*)_A$.  
Considering a single mode and converting to $S_n$ and $\phi_n$, the mean spontaneous emission and the zero-mean Langevin terms are
\begin{align}
R_n &=\frac{\Gamma}{\hbar\omega V_Av_g}2D_{nn} \\
  F_{Sn}&=\frac{\Gamma}{\hbar\omega V_Av_g}2 \Re(f_n^*F_n),&  F_{\phi n}&=\Im(F_n/f_n),
\end{align}

with correlations, \begin{align}
    \langle F_{Sn}(t)F_{Sn}(t')\rangle&=2S_nR_n\delta(t-t'),\\ \langle F_{\phi n}(t)F_{\phi n}(t')\rangle&=\frac{R_n}{2S_n}\delta(t-t').
\end{align}
With the diffusion constants established, we get the linewidth~\cite{coldren_diode_2012}
\begin{equation}
\Delta\nu=\frac{R_n(1+\overline{\alpha}^2)}{4\pi S_n}=\Delta\nu_{0} K_n|\Gammat_n|^2(1+\overline{\alpha}^2),
\end{equation}
where $\overline{\alpha}=\partial_N\Re(\omegan_n)/\partial_N\Im(\omegan_n)$ is the effective linewidth enhancement factor, and $\Delta\nu_{0}$ is the Schawlow-Townes linewidth for Fabry-Perot laser with $r_2=r_R(\omega_s)$. Inserting $\partial_N\omegan=\Gammat(i+\alpha)\Gamma v_gg_N$ in the expression for $\overline{\alpha}$ gives $\Delta\nu=\Delta\nu_{FP}/F^2$, where $\Delta \nu_{FP}=\Delta\nu_0K_n(1+\alpha^2)$ and $F$ is the chirp-reduction factor. 
\bibliography{references}
\end{document}